# Analytic performance comparison of routing protocols in master-slave PLC networks

Gerd Bumiller[1], Liping Lu[2] and YeQiong Song[2]

1 – iAd GmbH

Unterschlauersbacher-Hauptstr. 10,
D-90613 Großhabersdorf, Germany
Phone: +49 9105 9960-51

2 – LORIA – INPL – UHP Nancy 1

LORIA - TRIO
Campus Scientifique, B.P. 239
54506 Vandoeuvre, France
Phone : 33-(0)3-83 58 17 64

*Abstract*-In wide area master-slave PLC (powerline communication) systems, the source node cannot reach the destination node without packet relay. Due to the random channel characteristics in the powerline, the communication distance in terms of the number of repeaters between two nodes is not constant. So two kinds of dynamic repeater algorithms are developed: dynamic source routing and flooding-based routing. In this paper, we use analytic approach to compare the performance of those two routing protocols. We give formulas to calculate the average duration of a polling cycle for each protocol. Then we present simulation results to bolster the results of our analysis. Other metrics such as the bandwidth ratio occupied by routing protocol and the routing overhead within a packet have also been evaluated. Our results have shown that our flooding-based routing protocol always outperforms our dynamic source routing one in general.

I. INTRODUCTION

This work has been carried out as part of the REMPLI project (European program NNE5-2001-00825, www.rempli.org). The main objective of REMPLI is to develop a distributed infrastructure suitable for real-time monitoring and control of energy distribution and consumption.

The underlying communication system is based on a combined PLC and Internet network. The use of PLC technology is justified by the fact that on the one hand, the most of the energy metering and control equipments are already wired to the low-voltage and/or mid-voltage electric networks, and on the other hand, the difficulties to access to certain equipments using wireless technology as they are often located in closed environments with metallic obstacles (reinforced concrete walls and tubes). Moreover, as the geographic coverage of the applications that we should support largely exceeds the last-mile and in-house area networks, we only focus on the investigation of wide area PLC networks.

In a wide area PLC network, transmitting a packet from a source to a not immediately reachable destination node requires the packet relay of the intermediate nodes (repeaters). However, considering the dynamic topology change and impossible prediction of the powerline attenuation, repeaters cannot be statically configured. How to design the efficient routing protocols for dynamically adapting the powerline circumstances and shortening the transmission time under stringent bandwidth limitation consists in a challenging problem.

We further limit ourselves to the master-slave communication model for the following reasons. 1) This master-slave model matches well to the need of the monitoring and control applications since a controller naturally plays the role of the master to poll the meters (sensors) and to send the command to the actuators; 2) Master request and slave response scheme allows to efficiently avoid the packet collisions. Moreover the TDMA MAC protocol is assumed.

So our problem is further restricted to finding routing protocol for transmitting a packet from the master node to a slave node and a response (or confirm) packet in backward, and all this with the minimum transmission time.

As mentioned above, powerline channel exhibits random transmission characteristics. It is clear that the classic routing protocols like static source routing, dynamic distance vector or state link based routing used in Internet are not suitable.

Although Powerline channel has some similarity with the radio transmission channel, however routing protocols developed for wireless ad hoc networks [4] are inefficient since on the one hand, the master-slave model is not taken into account, and on the other hand, REMPLI PLC network has a relatively stable topology comparing to the mobile nodes in ad hoc networks.

For REMPLI PLC network, two routing protocols have been successively proposed: dynamic source routing of DLC1000 [1] and flooding-based routing over SFN [2]. In DLC1000, the master maintains the routing table by periodically getting information of the most interesting paths from slaves and determines the best routing path for transferring a packet to a slave. Corresponding repeater addresses are indicated in the packet header. A slave sends a packet back to the master using the reverse path. In SFN, all slaves can work as repeaters. The table of the number of repeaters (called hereafter repeater level) for reaching every slave is stored in the master. When a slave receives a packet correctly, it checks the packet header, and if the destination address is its own address or the remaining repeater level is zero, or the same packet is already transmitted once, it stops transmitting this packet, otherwise it continues to retransmit the packet by deceasing by one the repeater level.

For comparing these two protocols, we use an analytic approach to evaluate their performance for a set of common physical PLC topologies called "channel models". Each channel model is represented by a matrix of PER (Packet Error Rate), where PER[i,j] is the packet error rate of the transmission between two nodes i and j. In this paper we assume that PER is a time-constant value. In fact, according to [3] the interference in the PLC system, such as periodic impulse and noise, is considered as the Additive White Gaussian Noise, and consequently, the PLC network is considered as a non-time variant system. The result of the PER matrix is calculated by the physical layer emulator [3]. An example of PER matrix of *Np* nodes is show in Fig. 1. The first element of the matrix is defined as master, all others are slaves.

$$\begin{Bmatrix} 0 & \cdots & \cdots & PER_{M \to S_{Np-1}} \\ PER_{S_1 \to M} & \ddots & & PER_{S_1 \to S_{Np-1}} \\ \vdots & & \ddots & \vdots \\ PER_{S_{Np-1} \to M} & \cdots & \cdots & 0 \end{Bmatrix}$$

Fig. 1. PER matrix

The rest of this paper is organized as follows. In section II we define the comparison metrics. In sections III and IV, we analyze routing protocols of DLC1000 and SFN and give the formulas to evaluate their performance. In section V we compare on the one hand the analytic results with the simulation ones, and on the other hand the performance between the two routing protocols for a set of representative channel models. Finally, in section VI we give the conclusion and point out our future work.

## II. ROUTING PROTOCOLS AND PERFORMANCE METRICS

From the routing point of view, the routing efficiency is evaluated by the number of bits per data packet involving in the routing and the number of the control packets generated by routing protocols. Moreover additional packets for the reliable delivery should also be included. However, in REMPLI PLC network, the transmission time from master to each slave is a random variable, because of the different repeater levels for reaching each slave and the possible transmission errors leading to a random number of retries (although upper bounded) as well as the possible dynamic topology changes which lead to the changes of the repeater level of a path.

In order to compare the system performance of DLC1000 and SFN routing protocols, we focus on the following three performance metrics:

1) Average duration of a polling cycle: defined as the average time for the master to poll once all the slaves. For expressing the duration of a polling cycle exactly, we give the following formulas to calculate it for each protocols.
   - for DLC1000

$$D = \sum_{i=2}^{n} \sum_{j=0}^{n_{R\_i}} 2T_s \cdot \left( n_{R_j\_i} + 1 \right) \quad (1)$$

   parameters:
   n    number of node (master node with n=1)
   $T_s$    duration of one slot for transmitting a packet
   $n_{R\_i}$    retry number of node i
   $n_{Rj\_i}$    repeater level of node i for the $j_{th}$ retry

   - for SFN

$$D = \sum_{i=2}^{n} \sum_{j=0}^{n_{retry\_i}} \left( 2 \cdot (j+1) + r_{DL\_i} + r_{UL\_i} \right) \cdot T_s \quad (2)$$

   parameters:
   n    number of node (master node with n=1)
   $T_s$    duration of one slot for transmitting a packet
   $n_{retry\_i}$    retry number of node i
   $r_{DL\_i}$    repeater level of downlink (i.e., master to slave) of node i for the fist transmission
   $r_{UL\_i}$    repeater level of uplink of node i for the first transmission

2) Bandwidth consumed for routing signaling: calculated as the total number of routing information.
3) Routing overhead in a data packet: calculated as the number of routing information bits transmitted per data packet delivered.

## III. AVERAGE POLLING CYCLE DURATION OF DLC1000

It is the decision of the master to transmit without or with many repeaters and which slaves work as repeaters. The destination slave will send the packet back to the master according to the reverse path.

For transmitting a packet from the master to a slave and receive the confirm packet in backward without any repeater, 2 time slots are used and the probability of successful transmission from the master *M* to slave *s* is

$$\Pr\nolimits_{no\;retry,\;0\;rep.}(s,0) = (1 - PER_{M \to s}) \cdot (1 - PER_{s \to M}) \quad (3)$$

If the master decides to transmit with one repeater *R*, it will use 4 time slots and the probability of successful transmission from master *M* to slave *s* is:

$$\Pr\nolimits_{no\;retry,\;1\;rep.}\{PER_{M \to R_1}, PER_{R_1 \to S}\} = (1 - PER_{M \to R_1}) \cdot (1 - PER_{R_1 \to s}) \cdot (1 - PER_{s \to R_1}) \cdot (1 - PER_{R_1 \to M}) \quad (4)$$

For a fixed repeater level, there may be many paths from the master to the destination slave if there are many slaves between them. DLC1000 will decide a best path which has the highest successful transmission probability among paths. So the probability of successful transmission with one repeater level and using the best repeater $R_i$ is

$$\Pr\nolimits_{no\ retry}(s,1) = \max_{R_i \notin \{M,s\}} \left( \Pr\nolimits_{no\ retry,\ 1\ rep.} \{PER_{M \to R_i}, PER_{R_i \to s}\} \right) \quad (5)$$

The probability to get a successful transmission through a path of $n_R$ repeaters without retry is:

$$\Pr\nolimits_{no\ retry}(s,n_R) = \max_{R_i \notin \{M,s\}} \left( \begin{array}{l} (1-PER_{M \to R_1}) \cdot (1-PER_{R_1 \to R_2}) \cdots (1-PER_{R_n \to s}) \cdot \\ (1-PER_{s \to R_n}) \cdot (1-PER_{R_n \to R_{n-1}}) \cdots (1-PER_{R_1 \to M}) \end{array} \right) \quad (6)$$

The probability to get a successful transmission through a path of $n_R$ repeaters after $n^{th}$ retry is a random variable obeying the geometric distribution:

$$\Pr\nolimits_{n^{th}retry}(s,n_R) = \Pr\nolimits_{no\ retry}(s,n_R) \cdot \left(1 - \Pr\nolimits_{no\ retry}(s,n_R)\right)^n \quad (7)$$

The average timeslots for a success transmission with $n_R$ repeater is:

$$\bar{D}_{n_R}(s) = 2*(n_R+1) \cdot T_s \sum_{n=0}^{\infty}(n+1) \cdot \Pr\nolimits_{n^{th}retry}(s,n_R) \quad (8)$$

With some transformations the average duration $\bar{D}_{n_R}(s)$ is:

$$\bar{D}_{n_R}(s) = \begin{cases} \dfrac{2T_s \cdot (n_R+1)}{\Pr\nolimits_{no\ retry}(s,n_R)} & \text{for } 0 < \Pr\nolimits_{no\ retry}(s,n_R) \leq 1 \\ \infty & \text{for } \Pr\nolimits_{no\ retry}(s,n_R) = 0 \end{cases} \quad (9)$$

The minimum average duration of a polling cycle results in the maximum throughput of the network. The master is able to do this optimization and decides the number of repeaters. So the average duration of a polling cycle $\bar{D}_{rout,\Sigma}$ is the sum of the minimum average durations of all slaves.

$$\bar{D}_{rout,\Sigma} = \sum_{s \in S} \min_{0 \leq n_R \leq \max\{n_R\}} (\bar{D}_{n_R}(s)) \quad (10)$$

## IV. AVERAGE POLLING CYCLE DURATION OF SFN

We recall that in SFN, different repeater levels may be used in the downlink and the uplink, considering the powerline's random channel characteristics. The transmission of several nodes is independent. A same packet is only transmitted once per node.

Firstly, we analyze the case of downlink. Due to the flooding mechanism more than one participants transmit during the same time slot expect the beginning timeslot $t=1$. For repeater level $r$ ($=t-1$), every participant $\hat{s}$ ($\hat{s} \in \mathbf{S} = \{1...N_P\}, \hat{s}=1$ is the master) has the probability $\Pr\nolimits_{Tx,\hat{s}}(r)$ to transmit the packet. Time slot $t=1$ is defined, when the master transmit request ($\hat{s}=1$). The probability of transmission is then:

$$\Pr\nolimits_{Tx,\hat{s}}(0) = \begin{cases} 1 & \text{for } \hat{s}=1 \\ 0 & \text{else} \end{cases} \quad (11)$$

Then we calculate the correct reception probability $\Pr\nolimits_{Rcv,\hat{s}}(s,r)$ which is the probability of the first time correct reception of a packet in a time slot $t=r+1$.

If a slave does not receive a packet, it also does not transmit. If a slave has probability of transmitting a packet in this time slot, it necessarily means that the slave has correctly received the packet in the last time slot and has not already transmitted the packet before. So it exists that

$$\Pr\nolimits_{Tx,\hat{s}}(r) = \begin{cases} \Pr\nolimits_{Rcv,\hat{s}}(s,r-1) & \text{for } r=1 \\ \left(1 - \sum_{i=0}^{r-2}\Pr\nolimits_{Tx,\hat{s}}(i)\right) \cdot \Pr\nolimits_{Rcv,\hat{s}}(s,r-1) & \text{for } r>1 \end{cases} \quad (12)$$

We assume that the transmission of several transmitters is independent, and a receiver gets the packet from several transmitters. The case that a receiver does not receive a packet correctly is happened only when all transmitters fail to transmit the packet to the receiver. Otherwise, the receiver should have the probability to receive the packet correctly. So the first correct reception probability $\Pr\nolimits_{Rcv,\hat{s}}(s,r)$ is

$$\Pr\nolimits_{Rcv,\hat{s}}(s,r) = \left(1 - \sum_{v=0}^{r-1}\Pr\nolimits_{Rcv,\hat{s}}(s,v)\right) \cdot \left(1 - \prod_{s' \in \mathbf{S}|s' \neq \hat{s}} \left(1 - \Pr\nolimits_{Tx,s'}(r) \cdot (1 - PER(s',\hat{s}))\right)\right) \quad (13)$$

When we get the correct reception probability of each slave, the average duration for a downlink transmission per slave can be calculated. In this case, we should consider the retry in the calculation. For a defined repeater level, if the correct reception probability is not equal to 1, there exits the probability of transmission failure. When a transmission fails, the retry happens and the master should increment by one its repeater level.

The process of retry used in SFN protocol can be modeled as a Markov chain. The proof is below.

**Proof:** Let $R_t$ represent the repeater level of the network at time t. Note that each transmission is independent. If using the current repeater level, the transmission fails, the repeater level will be incremented by one and the master enters the next repeater level. So we show that the random process $\{R_t\}$ has the Markovian property, i.e., the future behavior of the random process depends on the current state only, and not on the sequence of states it took to reach the current state.

Also, if the transmission successes, the repeater level will be the actual repeater level used in this transmission. So we construct the Markov chain as shown in Fig. 2.

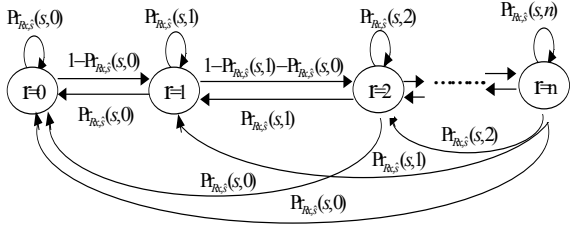

Fig. 2. Markov chain model of repeater level

Let $\pi^t = \{\pi^t(0), \pi^t(1), \pi^t(n)\}$ denote the state probability vector of the Markov chain at time $t$. The state probability vector $\pi^t$ converges to a stationary steady-state vector $\pi$, i.e., $\lim_{x \to \infty} \pi^t = \pi$. When we get the repeater level stationary steady-state vector $\pi$, the average repeater level for a downlink transmission per slave and the average over all slaves can be obtained by:

$$\overline{R}_{DL}(s) = \sum_{r=0}^{\max\{r\}} r \cdot \pi(r) \quad (14)$$

If the number of allowed repeater levels $\max\{r\}$ is not enough, we calculate equations as above and terminate the calculations at $r > \max\{r\}$. The useful values of the allowed number of repeater level for the downlink is:

$$r_{DL}(s) \in \left\{ \lfloor \overline{R}_{DL}(s) \rfloor ; \lceil \overline{R}_{DL}(s) \rceil \right\} \quad (15)$$

For the uplink, after receiving a packet correctly, the destination slave $s$ has the transmission probability $\Pr_{Tx,\hat{s}}(0)$ to transmit the response/confirm back to the master.

$$\Pr_{Tx,\hat{s}}(0) = \begin{cases} \sum_{r=0}^{r_{DL}(s)} \Pr_{Rcv,1}(s,r) & \text{for} \quad \hat{s} = s \\ 0 & \text{else} \end{cases} \quad (16)$$

And the probability that the master correctly receive the response/confirm denoted by $\Pr_{Rcv,s}(1,r)$ can be evaluated in a similar way of the downlink. The useful values for the allowed number of repeater levels for the up link are:

$$r_{UL}(s) \in \left\{ \lfloor \overline{R}_{UL}(s) \rfloor ; \lceil \overline{R}_{UL}(s) \rceil \right\} \quad (17)$$

Next, we will analyse the average duration of a transmission including a downlink and an uplink. The master has to decide the allowed number of repeater levels for the downlink $r_{DL}(s)$ and for the up link $r_{UL}(s)$. Even if the destination slave receives the request before the allowed number of repeater levels is finished, it is not possible to send the response, without a strong danger of collision. So the probability of successful polling slave $s$ with average repeater levels of downlink and uplink and without retry is:

$$\Pr(s) = \left( \sum_{r=0}^{r_{DL}(s)} \Pr_{Rcv,1}(s,r) \right) \cdot \left( \sum_{r=0}^{r_{UL}(s)} \Pr_{Rcv,s}(1,r) \right) \quad (18)$$

The probability of successful polling slave $s$ with average repeater levels of downlink and uplink after $n^{th}$ retry is:

$$\Pr_{n^{th} \, retry}(s) = \Pr(s) \cdot (1 - \Pr(s))^n \quad (19)$$

The average transmitted number (retry number+1) is:

$$E(n+1) = \sum_{n=0}^{\infty} (n+1) \cdot \Pr(s) \cdot (1 - \Pr(s))^n \quad (20)$$

The average duration of a transmission task $\overline{D}_{SFN}(s)$ for one slave is equal to the average transmission time multiplied by the average transmitted number.

$$\overline{D}_{SFN}(s) = T_s \cdot (2 + r_{DL}(s) + r_{UL}(s)) \cdot \sum_{n=0}^{\infty} (n+1) \cdot \Pr(s) \cdot (1 - \Pr(s))^n \quad (21)$$

With some replacements it is possible to show, that it is a binomial series with negative exponent and the result is:

$$\overline{D}_{SFN}(s) = \frac{(2 + r_{DL}(s) + r_{UL}(s)) \cdot T_s}{\Pr(s)} \quad \text{for } \Pr(s) > 0 \quad (22)$$

We use all allowed numbers of repeater levels in (15) and (17) to calculate respective value to find minimum $\overline{D}_{SFN}(s)$.

The average duration of a polling cycle is:

$$\overline{D}_{SFN,\Sigma} = \sum_{s \in \mathbf{S} | s \neq 1} \min\{\overline{D}_{SFN}(s)\} \quad (23)$$

V. COMPARISON

For the analysis, 5 channel models are used, which are Ring_10, Ring_100, Rand_Area Np_20, Rand_Area Np_100, Rand_Area Np_200. The two former channel models have the topology of ring and the latter three have the topology of a tree with the master as the root and the randomly distributed slaves as leaves. The number in the channel model name indicates the number of the nodes.

*A. Comparing analytic results with the simulation ones*

The maximum repeater level is different in DLC1000 and SFN. In DLC1000, the maximum repeater number is chosen by the master for the best performance of the system. In SFN, because after each retry, the repeater level increments by one, the maximum repeater level is the repeater level at which no participant has any probability to transmit a packet and the transmission has finished.

In TABLE I, we compare the calculated values of average duration of a polling cycle with simulation results [5]. In the implementation of DLC1000, the maximum repeater level is defined as two and only two repeater addresses are defined in the packet header. So in some channel models, with two repeater levels, the master cannot reach some slaves. Nevertheless simulation results are still reported by precising the number of reached nodes.

From TABLE I and II, we see that the analytic results are close to the simulation results. And we also notice that there is a big difference between the analytic result and the simulation one in TABLE II for the case of RandArea_20 channel model. This is due to the particularity of this channel model and the simulation program. In channel model Rand_Area Np_20, some slaves cannot be connected by the master at some intervals. In SFN simulation, the retry number is configured to two. After two retries, the master considers the slave as disconnected. The master will then try to reconnect the slave. It is so the network management task's time that is included into our simulation results. In contrary, in DLC 1000 simulation, if the master finds that one slave cannot be reached after the maximum retry number, the simulator will stop the polling, and go back to initial phase, guarantee all slaves in the connect state, then the polling cycle restarts.

TABLE I
Analytic and simulation results of DLC1000

| Channel Model | Maximum repeater number | $\overline{D}_{rout,\Sigma}$ | Simulation results | Relative difference |
|---|---|---|---|---|
| Ring_10 | 2 | 30.0 | 30.94 | -3% |
| Ring_100 | 3 | 427.3 | 399.12 (91 slaves) | |
| RandArea Np_20 | 2 | 63.0 | 76.3 | -17.4% |
| RandArea Np_100 | 3 | 420.6 | 403.68 (90 slaves) | |
| RandArea Np_200 | 4 | 1026.6 | 806.03 (160 slaves) | |

TABLE II
Analytic and simulation results of SFN

| Channel Model | Maximum repeater number | $\overline{D}_{SFN,\Sigma}$ | Simulation results | Relative difference |
|---|---|---|---|---|
| Ring_10 | 4 | 28.4 | 30.6 | -7.2% |
| Ring_100 | 3 | 419.1 | 429.9 | -2.5% |
| RandArea Np_20 | 3 | 42.2 | 90.8 | -53.5% |
| RandArea Np_100 | 7 | 380.5 | 419.3 | -9.3% |
| RandArea Np_200 | 6 | 987.1 | 1033.9 | -3.0% |

*B. Average duration of a polling cycle*

The following table compares the results of the theoretical analysis of DLC1000 and SFN.

TABLE III
Average duration of SFN and DLC1000

| Channel Model | SFN | DLC 1000 |
|---|---|---|
| Ring_10 | 28.4 | 30.0 |
| Ring_100 | 418.3 | 427.3 |
| RandArea Np_20 | 41.4 | 63.0 |
| RandArea Np_100 | 379.9 | 420.6 |
| RandArea Np_200 | 983.4 | 1026.6 |

Even with optimal assumptions for DLC1000 routing, the performance of SFN is the best in all analysed channel models.

*C. Bandwidth consumed for routing signaling*

In the initial phase, the master works in the similar ways in two protocols. The master sends packets to try to connect with all the slave using the serial number and distribute a network address for each connected slave. After the initial phase, the master will periodically send the packet to request slaves which are in the state of connection or not. There is a difference concerning the slave responses. For SFN, only the confirmation will be send back. For DLC1000, besides the confirmation in the response packet, the slave will add five preferable repeaters and the corresponding evaluated channel quality which is the routing information used to decide the best routing path by the master. So obviously, the bandwidth consumed for routing signalling in DLC1000 is more important than in SFN.

*D. Routing overhead*

In the same packet length, for example 64 bytes packet, the packet headers of two protocols are different about the routing information.

TABLE IV
Routing overhead

| | Routing information (bits) | Routing bits / total packet |
|---|---|---|
| DLC1000 | Two repeater address (2*12bits) | 4.7% |
| SFN | Repeater levels (8 bits) 4bits for downlink 4bits for uplink | 1.6% |

VI. CONCLUSION

From the above comparison, we can withdraw that in general case, the flooding-based routing protocol of SFN outperforms the dynamic source routing protocol of DLC1000. In fact, for a given receiver, there are more transmitters in SFN than in DLC1000 for the repeater level more than one. Because the transmission of each transmitter is independent, in SFN not only the best path is used but also all other possible paths, the correct reception probability is therefore higher in SFN than when only a single transmitter is used (case of DLC1000). The analytic results and simulation results prove this conclusion. For the future work, we will extend our formulas for studying the performance of the two protocols in the transient phase such as the initial phase and in the special topologies such as a ring topology with random switch open (where a channel becomes time-variant).